\documentclass[11pt,a4paper,aps,showpacs,superscriptaddress,nofootinbib]{revtex4-1}

\usepackage{epsfig}
\usepackage{float}
\RequirePackage{amssymb,amsmath}
\usepackage[latin1]{inputenc}
\usepackage{graphicx}
\usepackage{indentfirst}
\usepackage{color}
\begin{document}

\title{Investigation of $^{11}$B+$^{197}$Au Reaction at Intermediate Energies}

\author{G. S. Karapetyan}
\affiliation{CCNH,  Universidade Federal do ABC
 09210-580, Santo Andre, Brazil}
\email{gayane.karapetyan@ufabc.edu.br}

\author{N. A. Demekhina}
\affiliation{Yerevan Physics Institute, Alikhanyan Brothers 2,
Yerevan 0036, Armenia}
\affiliation{Joint
Institute for Nuclear Research (JINR),  Laboratory of Nuclear Problems
(LNP), Joliot-Curie 6, Dubna 141980, Moscow, Russia}
\email{demekhina@nrmail.jinr.ru}

\author{A. R. Balabekyan}
\affiliation{Yerevan State University,
A. Manoogian, 1, 025, Yerevan, Armenia}
\email{balabekyan@ysu.am}

\begin{abstract}
Mechanism of nuclear reactions on $^{197}$Au induced by $^{11}$B ions at energies above Coulomb barrier was studied by induced-activity method and $\gamma$-spectroscopy. 
The cross sections of the reaction fragments from $^{197}$Au induced by $^{11}$B ions were measured at bombarding energies 137.5 and 255.5 MeV. 
The fission process was investigated by using multimodal fission approach at the energy 137.5 MeV, and
pure symmetric distribution at 255.5 MeV.
It was observed that the transferred linear momentum provides the information on the initial projectile-target information. The fissility for both fission reactions under study was deduced from measured fission cross section using the total inelastic cross section. Comparison with proton-induced fission shown, that the linear momentum transferred to the fissile system depends on the probe.
\end{abstract}

\maketitle

\section*{1 Introduction}

The study of nucleon-nucleus and nucleus-nucleus collisions is a source of experimental data, which are extremely important for scientific and technological application. There are multiple questions addressed in the context of e.g., design of accelerator driven systems of energy amplification and nuclear waste utilization, astrophysical studies, radiological safety, etc. which cannot be answered without knowledge of collision cross section.
A typical characteristics of the nuclear structure of interacting nuclei can be revealed at energies near the Coulomb barrier. Particularly, the study of reaction mechanism concerning complete fusion (CF) and incomplete fusion (ICF), as well as the transition between them is important in the view of understanding the interplay between such two dominant modes of the nuclear interaction.

In the frame of the dynamic model \cite{Wilczynski}, during the amalgamation of the projectile with target nucleus the complete-fusion cross section reaches a maximum at the energy for which
the angular momentum, imparted to the target in the entrance reaction channel, lower than some critical value, $l_{crit}$. According to the model, above this critical momentum the collision become more peripheral and compound system cannot exist. With increasing the incident energy, incomplete fusion channels, when the different parts of the projectile can captured by the target, become more probable.
Among the reaction products there are some, which are originated due to two different mechanisms: projectile break-up and nucleon coalescence during the thermalization of the composite nuclei
produced in the complete and incomplete fusion \cite{Gadioli1, Gadioli2}. This processes are assumed to be a peripheral and at considerable low excitation energy in the course of the interaction target and projectile. In the work \cite{Aguilera} it was observed the increasing of ICF contribution at high energies and angular momentum, imparted to the system, exceeding the limit value for fusion process what is more probable in direct surface interaction. On the other side, there are suggestions that there is no sharp boundary in the coexistence of the CF and ICF processes. Such coexistence is possible at both low and high energies just with different proportion. The reactions, proceeding in the frames of statistical presentation and describing by statistical models, usually are related to CF data.  As a rule, the disagreement of the model calculations of compound nucleus decay without the projectile breakup with the experimental results can be indicated as ICF contribution.

$^{11}$B nuclei can be presented as stable nucleus having comparatively low binding energy relative to decay into separated clusters according to the following schemes:  

  $^{11}B\longrightarrow^{7}Li + ^{4}He$    with  Q = -8.665 MeV;   
	
  $^{11}B\longrightarrow^{8}Be + ^{3}H\longrightarrow$     with Q = -11.224 MeV
	
                  $\longrightarrow^{4}He + ^{4}He + ^{3}H$     with Q = -11.132 MeV.

A capture of $^{11}$B or one of above-mentioned components of $^{11}$B would form the composite nuclei during complete or incomplete fusion. Following the evaporation, the pre-equilibrium processes and the sequence $\alpha$- or/and $\beta^{\pm}$ - decays do not allow to separate the initial step of the fusion. The  main channels of the statistical decay  can be represented by the following chain:   
			
Fission ($f$) + $\alpha$-emission ($\alpha$) + proton emission (p) + neutrons emission (xn). 

These decay channels can be combined in different ways which depend on the excitation energy, transferred momentum and nuclear properties of the nuclei formed in reaction. 

Among whole processes of nucleus-nucleus interaction, fission represents the most interesting phenomena. Fission is a slow process on a nuclear timescale, involving deformation of the whole nucleus. Studies of the mass and charge distributions of the fission fragments at intermediate energies provide important information about the dynamics of the reaction. Determination of cross sections for the interaction of charged particles and heavy ions with nuclei reveals that the reaction mechanism for compound system formation varies with incident particle energy. The dependence of experimental data of fission at medium excitation energies show that at the time of formation of the charge, mass and energy distributions an essential role is played angular momentum of the fissioning nucleus transferred in the entrance channel of the reaction. Thus the process of scission and formation of fragments is not only influenced by the temperature, but also by the total angular momentum of the fissioning nucleus.

While fission of actinides has been investigated in details using different projectiles as photons, protons, heavy-ions in the large energetic scale, fission of of pre-actinide nuclei is limited. A considerable amount of measurements have been performed for proton-induced reactions on the gold target using various experimental methods and technique and predicted by theoretical models, were compiled and analyzed in the well known work of A. V. Prokofiev \cite{Prokofiev}. Experimental data for the light and heavy-ion-induced fission of gold are not rich enough and presented mainly by some works with $^{4}$He and $^{12}$C projectiles \cite{Buttkewitz, Gindler, Lisman}. However, no literature data exists on the yield distribution of fission product, in the reaction of $^{11}$B gold target neither at low nor at intermediate energy range.

To avoid the lack of data, the experiment, which the subject is $^{11}$B-induced reaction on gold target at intermediate energies was performed. It is of particular interest to study also the variation fission cross section and fissility with angular momentum transferred in the entrance channel of reaction for nucleon-and heavy-ion-induced fission.
In addition, the charge and mass distributions of the fission fragments at 137.5 and 155.5 MeV were used to determine the fission cross section. On the base of the dynamic model, based on the force equilibrium concept \cite{Wilczynski}, it was possible to extract the angular momentum imparted to the fissioning nucleus at initial stage of the interaction.

\section*{2 Experimental Procedure}

The target consisting of six natural gold foils of 10 $\mu$m thickness each one 
were sandwiched by six Al foils of 47 $\mu$m thick each.
All foils were piled up together and aligned perpendicular to the
beam direction, were exposed to an accelerated
$^{11}$B-ion beam of initial energy 23.6 MeV/n from the LNR Phasotron, Joint Institute for
Nuclear Research (JINR), Dubna, Russia.
Aluminium degrading foils were used in order to obtain a reduction in the beam energy, which average values at the center of Au foils were 255.5 and 137.5 MeV.
The irradiation time was 16 hours at ion beam intensity of about 7.8 $\times$ 10$^{13}$ nuclei per hour.
The measurements of the spectra of $\gamma$-rays emitted in the decays of radioactive reaction products began 10 min after the completion of the irradiation and lasted five months by using HpGe detector with energy resolution 0.23\% at an energy of 1332 keV. The energy-dependent detection efficiency of the HpGe detector was measured with standard calibration sources of $^{22}$Na, $^{54}$Mn, $^{57;60}$Co, $^{137}$Cs,  $^{154}$Eu, $^{152}$Eu, and $^{133}$Ba. The half-lives of identified isotopes were within the range between 15 min and 1 yr. The error in determining cross sections depended on the following factors: the statistical significance of experimental results ($\leq$ 2-3\%), the accuracy in measuring the target thickness and the accuracy of tabular data on nuclear constants ($\leq$ 3\%), and the errors in determining the detector efficiency with allowance for the accuracy in calculating its energy dependence ($\leq$ 10\%).

The fragment production cross sections are usually considered as an
independent yield (I) in the absence of a parent isotope (which may
give a contribution in measured cross section via $\beta^{\pm}$-
decays) and are determined by using the following equation:

\begin{eqnarray}
\hspace{-0.2cm}\sigma=\frac{\Delta{N}\;\lambda}{N_{d}\,N_{n}\,k\,\epsilon\,\eta\,(1-\exp{(-\lambda
t_{1})})\exp{(-\lambda t_{2})}(1-\exp{(-\lambda t_{3})})}\label{g1}
\end{eqnarray}
\noindent where $\sigma$ is the cross section of the reaction
fragment production (mb); $\Delta{N}$ is the area under the
photopeak; $N_{d}$ is the deuteron beam intensity (min$^{-1}$);
$N_{n}$ is the number of target nuclei (in 1/cm$^{2}$ units);
$t_{1}$ is the irradiation time; $t_{2}$ is the time of exposure
between the end of the irradiation and the beginning of the
measurement; $t_{3}$ is the measurement time; $\lambda$ is the decay
constant (min$^{-1}$); $\eta$ is  the intensity of
$\gamma$-transitions; $k$ is the total coefficient of $\gamma$-ray
absorption in target and detector materials, and $\epsilon$ is the
$\gamma$-ray detection efficiency.

In the case where the cross section of a given isotope includes a
contribution from the $\beta^{\pm}$-decay of neighboring unstable
isobars, the cross section calculation becomes more complicated
\cite{Gaya}. If the formation cross section of the parent isotope is
known from experimental data, or if it can be estimated on the basis
of other sources, the independent cross sections of daughter nuclei
can be calculated by the relation:

\begin{widetext}
\begin{eqnarray}
\sigma_{B}=&&\frac{\lambda_{B}}{(1-\exp{(-\lambda_{B}t_{1})})\,
\exp{(-\lambda_{B}t_{2})}(\,1-\exp{(-\lambda_{B} t_{3})})}\times\nonumber\\&&
\hspace*{-1.5cm}
\left.\Biggl[\frac{\Delta{N}}{N_{\gamma}\,N_{n}\,k\,\epsilon\,\eta}-
\sigma_{A}\,f_{AB}\,\frac{\lambda_{A}\,\lambda_{B}}{\lambda_{B}-\lambda_{A}}
\Biggl(\frac{(1-\exp{(-\lambda_{A} t_{1})})\,\exp{(-\lambda_{A} t_{2})}\,(1-
\exp{(-\lambda_{A} t_{3})})}{\lambda^{2}_{A}}\right.\nonumber\\
&&\left.\qquad
-\frac{(1-\exp{(-\lambda_{B} t_{1})})\,\exp{(-\lambda_{B}
t_{2})}\,(1-\exp{(-\lambda_{B} t_{3})})}{\lambda^{2}_{B}}\Biggr)\right.\Biggr],
\end{eqnarray}
\end{widetext}
\noindent where the subscripts $A$ and $B$ label variables referring
to, respectively, the parent and the daughter nucleus; the
coefficient $f_{AB}$ specifies the fraction of $A$ nuclei decaying
to a $B$ nucleus ($f_{AB}=1$, when the contribution from the
$\beta$-decay corresponds 100\%); and $(\Delta{N})_{AB}$ is the
total photopeak area associated with the decays of the daughter and
parent isotopes. The effect of the precursor can be negligible in
some limiting cases: where the half-life of the parent nucleus is
very long, or in the case where its contribution is very small. In
the case when parent and daughter isotopes can not be separated
experimentally, the calculated cross sections are classified as
cumulative ones (C). It should be mentioned that the use of
induced-activity method imposes several restrictions on the
registration of the reaction products. For example, it is impossible
to measure a stable and very short-lived, very long-lived isotopes.

\section*{3 Results and Discussion}

In the reaction induced by 137.5 MeV and 255.5 MeV $^{11}$B ions on $^{197}$Au target, the production cross sections were determined for 96 and 107 target fragments, respectively, in the mass range of $7 \leq A \leq 205$ u. These data are summarized in Table I. 

In order to obtain a complete picture of the charge and mass distributions of reaction products is necessary to estimate the cross sections of isotopes unmeasurable by the activation method. It is necessary, therefore, to estimate the charge distribution curve (i. e., the variation of cross section with $Z$ at constant $A$) using independent cross section of the reaction products. Such variation can usually be expressed as a Gaussian distribution function \cite{Kudo}:
\begin{eqnarray}
\sigma_{A,Z}=\frac{\sigma_{A}}{(C\pi)^{1/2}}exp({-\frac{(Z-Z_{p})^{2}}{C}}),
\end{eqnarray}
\noindent where $\sigma_{A, Z}$ is the independent cross section for a given nuclide with an atomic charge $Z$ and a mass number $A$, $\sigma_{A}$ is the total isobaric cross section of the mass chain $A$, $Z_{p}$ is the most probable charge for that isobar, and $C$ is the width parameter of the distribution for the mass number $A$. Parameters of charge distribution determine the position of residue nucleus concerning stable isotopes with maximum yield in isobaric chain.

In the assumption of the constant width parameter of charge distributions ($C$) for different mass chains \cite{Kudo, Branquihno}, least-squares method was applied in order to get fitting parameters $Z_{p}$ and $\sigma_{A}$. The cross section of a particular isotope ($Z, A$) may be independent or partly or completely cumulative, depending on decay chains of precursors. 

Given the assumption of Gaussian charge distribution, the beta-decay feeding correction factors for cumulative yield isobaric members can be calculated once the centroid and width of the Gaussian are known. In order to uniquely specify the variables $\sigma_{A}$ and $Z_{p}$ one would need to measure three independent cross sections for each isobar. In fact, the nature of radioanalytical studies such as this one does not, in general, lend itself to the measurement of isobaric members. Rather, a wide assortment of radioactivities are observed which span the entire range of the periodic table that is accessible in the nuclear reaction. As a result relatively few isobaric pairs are
observed. Another assumption is that the charge distribution curves for neighboring isobaric chains should be similar; thus radionuclide cross sections from a limited mass range can be used to determine a single charge distribution curve.

In such a way the measured production cross sections are adjusted to remove precursor feeding, where necessary, and a set of independent cross sections were generated. The calculated values of $\sigma_{A}$ that correspond to the total isobaric cross section for a specific mass number made it possible to construct the mass-yield distribution. The mass-yield distribution determined for the reaction induced by 137.5 and 255.5 MeV $^{11}$B ions on the $^{197}$Au targets are present in Figs. 1 and 2, respectively,
with inclusion the fission product range, spallation and the light fragment production on either side.
From the Figs. 1 and 2 one can see the clear-cut distinction between spallation or deep spallation and fission mass ranges, which represents by nearly symmetric peak in the mass yield curve at about half the target mass of gold.

The heavy residual nuclei measured in the present work were produced in reactions induced with $^{11}$B at sufficiently high energies. Significant fraction of the reaction cross sections in this case is stipulated for higher angular momenta, where the complete fusion is prevented by centrifugal forces and the break-up processes accompanied by emission of the projectile fragments start to play an increasing role. Correspondingly, the contribution of the incomplete fusion increases. 
The survivals  of the compound nucleus at the high excitation energies may be define in the comparison with the model considering the compound nucleus as a statistical equilibrium system undergoing de-excitation through emission light particles and fission.

As a result of the interaction, the heavy residual products near target mass number can be formed in different reaction channels, as a deep inelastic scattering accompanying by emission nucleons and light particles (DIS) (including direct transfer processes), complete fusion - evaporation (CFEP), complete fusion - fission (CFF), and incomplete fusion - evaporation (ICFEP).  In present experiment the complete fusion leads to the formation of $^{208}$Po compound nucleus. In the case of incomplete fusion the composite nuclei as $^{205}$Bi, $^{204}$Pb and $^{201}$Tl could be formed.
We suggest that isotopes, Po, Bi, Pb, Po and Os, which originate in the range close to the target mass $(\Delta A = A_{CN} - A) \leq 20$ u, where $A_{CN}$ is a mass of compound nucleus, are heavy residues, which form during CFEP and ICFEP processes. As one can see from Table I, the main part of the measured cross sections represent the cumulative yields and its excitation functions could be result of both CFEP and ICFEP processes. As it seen from Table, isotopes in mass range 196-198 u, namely, near target mass number, are produced with high probability. These residuals can be formed via different processes, in which just a few nucleon change between target and projectile, such as inelastic scattering and different direct reactions on the surface of the target \cite{Gadioli1, Gasques, Eyal}. 

The region of light isobars would correspond to the light fragment production
One of the possible mechanism for these fragments is that they would correspond to the counterpart pair of products in the mass region $A \sim 110-120$. Also they could be originated from deep spallation process which nuclides would emit not only
nucleons and light charge particles (with $Z \leq 2$) but also some heavier elements in the IMFs region. Moreover, an alternative explanation of the origin of light fragments was suggested
by the intranuclear cascade model \cite{Yariv}, according to which these fragments are the result of the disintegration of highly excited residual nucleus with $A \sim 185$.

The important region in mass-yield distribution would correspond to the wide distributions at mass ranges $A = 55-145$ at both energies. The centroids of those distributions suggest that these products may be result of binary fission of target-like species. 
The mass and energy distributions of fragments in the fission of nuclei from Pb to No 
\cite{Gonnenwein} have confirmed the validity of a hypothesis concerning the existence of independent fission modes, as first stated by Turkevich and Niday \cite{Turkevich}.
This hypothesis has received physical substantiation in theoretical works by Pashkevich \cite{Pashkevich} and Brosa $\textit{et al.}$ \cite{Brosa}. These studies have shown that multimodality of the mass and energy distributions of fission fragments is caused by the valley structure of the deformation potential energy surface of a fissioning nucleus. The experimental mass and energy distributions from the fission of actinide nuclei are usually assumed to consist of different mass and energy distributions for two independent fission modes: symmetric (S) and asymmetric (AS). Mode S is mainly conditioned by the liquid drop properties of nuclear matter, and therefore the most probable values of fragment masses $A$ are close to $A_{f}/2$, where $A_{f}$ is the mass of the fissioning nucleus. The asymmetric mode AS with average masses of the heavy and light fragments $A_{H}$, $A_{L}$ with $Z_{H}$, $N_{H}$ and $Z_{L}$, $N_{L}$ (proton and neutron numbers of a heavy and light fragments) close to the one of the shell numbers.

In the case of $^{197}$Au target at 137.5 MeV, the fission cross section as a function of mass number is obtained by the sum of three Gaussian functions, corresponding to symmetric and asymmetric fission modes \cite{Younes}: 
\begin{align}
 \begin{split}
\sigma_{A} = &
\frac{1}{\sqrt{2\pi}}\bigg[\frac{K_{AS}}{\sigma_{AS}}
\exp\left({-\frac{(A-A_S-D_{AS})^2}{2\sigma^2_{AS}}}\right)+
\frac{K'_{AS}}{\sigma'_{AS}}\exp\left(-\frac{(A-A_S+D_{AS})^2}
{2\sigma'^2_{AS}}\right)+\\
 & \frac{K_S}{\sigma_S}\exp\left({-\frac{(A-A_S)^2}
{2\sigma^2_S}}\right)\bigg],
\label{mass}
 \end{split}
\end{align}
\noindent where $A$ is the fragment mass number; $\overline{A}_{S}$ is the mean mass number which determines the center of the Gaussian functions; and $K_{i}$, $\sigma_{i}$, and D$_{i}$ are the contribution, dispersion and position parameters of the $i^{th}$ Gaussian functions. The indexes $AS$ and $S$ designate the asymmetric and symmetric components.

The total reaction cross sections at both energies were calculated by summing the cross section for spallation, light nuclide production and fission cross section (Table II).
The results of the fitting procedure at 255.5 MeV considered only the symmetric fission component of the mass-yield distributions can be seen on Fig. 2.
Integrating over the Gaussian and multiplying with a factor 0.5, because of the two fission fragments in each fission event, gives an estimate for the fission cross section. In Figs. 1 and 2 the mass-yield distributions for 137.5 MeV and 255.5 MeV, obtained by fitting procedure, is represented by the solid curves. 

The values of the fit parameters together with the fission cross section are tabulated in Table II. Analysis of the mass distribution curves made it possible to determine the positions of peak and the width of mass distributions. 
One can see in Table II, the value of the width at 137.5 MeV of the present work is in good agreement with data for $\alpha$-induced fission of gold at 140 MeV of incident particle \cite{Buttkewitz}.

One can see from Fig.1 that at study of mass-yield distributions of fission fragments obtained in the reaction $^{197}$Au+$^{11}$B at energy 137.5 MeV leading to the formation of highly excited composite systems, the asymmetric fission mode is evident, manifesting itself in the form of narrow ``shoulders". 

Thus, we have an indication that the shell structure of the fragments formed in the ranges of the light $A_{L}=59$ u for 137.5. MeV comprises spherical light fragment with $N\sim50$, and heavy $A_{H}=139$ u for 137.5. MeV, influenced by the deformed neutron shell closure $N=88$, respectively, favor the fission process, the shells in both light and heavy fragments still playing definite role and not vanish completely. 

The values of the fit parameters, tabulated in Table II, show a maximum around mass $A = 99$ and 97, which corresponds the masses $A = 198$ and 194 for the fissioning nuclei at lower and higher energies, respectively. The estimated part of the fission cross section originating from asymmetric fission mode were found to be 0.3 \% at 137.5 MeV. One can see from Fig.2 that the contribution of asymmetric fission to the total fission yield decreased up to zero. Thus, we can conclude the shell effects are gradually smeared at vanished completely at intermediate energy regime.

From the mean values of the mass distributions it can be concluded that on the average $\sim$ 10 neutrons and $\sim$ 14 are emitted before and after fission at low energy and at high energy of incident ions, respectively. Therefore a masses of 198 u and 194 u are expected for the fissioning nuclei in contrast with mass $A = 208$ of the compound nucleus. 
The average number of emitted neutrons at 137.5 MeV for B-induced fission of the present work is in good agreement with value 11.1 for $\alpha$-induced fission \cite{Buttkewitz}, giving a clear indication that it is not the incident particle but the structure of the fissioning nucleus and the characteristics of the fragments which determine the mass distribution.

The mass-yield distributions of fragments depend on the mass of the fissioning nuclei and on its excitation energy. During fitting procedure it was found that as the excitation energy increases, the distribution becomes slightly wider. It is also worth noting that the values of mean mass number of distributions after evaporation of neutrons $M_{A}$ at 255.5 MeV is shifted to lower mass range in comparison to the 137.5 MeV. It means, that with the increasing of excitation energy, the number of evaporated neutrons from the fissioning nuclei and the fission fragments is increased too.

Neutron evaporation is a process that competes with the fission process \cite{Duijvestijn, Maslov}. The high excitation energy is coupled to large number of the neutron evaporation. In the case of interactions with high-energy particles, the fission process is considered at a slow reaction stage after the completion of the internuclear cascade and the formation of a set neutron-deficient fissioning nuclei with higher fissility parameter. 

The total fission cross sections in this work are presented in Table II.
This values exceed the fission cross sections for $\alpha$-induced fission at almost the same incident energy by a factor $\sim$ 5-6. At the analysis of fission induced by the accelerated heavy ions, the effect associated with transferred angular momentum was obtained.

As it was shown from \cite{Sierk, Wilczynski}, the increase in rotational energy affects on fission possibility. In the energy range studied in this paper, we assume a linear dependence of the transferred angular momentum from the mass of the projectile. According to the dynamic model \cite{Wilczynski} heavy-ion induced fusion-fission reactions are characterized by the formation of a fully equilibrated compound nucleus where the initial relative kinetic energy and angular momentum of the projectile is converted into the intrinsic excitation energy and spin of the fused system. Calculated fission-barrier heights as a function of angular momentum have shown a lowering of the fission barriers with increasing of angular momentum from zero value. It can be seen that fission is expected to play a significant role only above 50  $\hslash$, when the fission barrier has dropped to about half of the value it has at zero angular momentum.

At energies above the barrier, the formula for calculation for the average angular momentum, $< \ell >$ is provided by \cite{Capurro}:
\begin{eqnarray}
< \ell >=\frac{2}{3}\sqrt{\frac{2\mu R^{2}(E_{c.m.}-V_{CB})}{\hslash^{2}}}.  \label{9}                                                                       
\end{eqnarray} 
\noindent
$R$ represents the impact parameter of the collision and can be calculated on the basis of experimental total reaction cross section, using the hard sphere model \cite{Bradt} for nucleus-nuclear interactions:
\begin{eqnarray}
\sigma_{tot}=\pi r^{2}_{0}(A^{1/3}_{T}+A^{1/3}_{p}-b_{Tp})^{2}    fm^{2},
\end{eqnarray}
\noindent where $A_{T}$ and $A_{p}$ are the mass numbers of the target and projectile nuclei, respectively; $r_{0}$, is the constant of
proportionality in the expression of geometrical nuclear radius $r_{i}=r_{0}A^{1/3}_{i}$ and $b_{Tp}$ is the overlap parameter. The value of $< \ell >$ calculated by (\ref{9}) for $^{197}$Au at 137.5 and 255.5 MeV were $< \ell >  = 47\pm$5 and 72$\pm$7 $\hslash$, respectively.

Putting the value of the experimentally determined total reaction cross section in expression (6) and calculate the value for $b_{Tp}$,
the impact parameter $R$ can be estimated using relation:

\begin{eqnarray}
R = r_{0}(A^{1/3}_{T}+A^{1/3}_{p}-b_{Tp})    fm.
\end{eqnarray}

Using the obtained values of average angular momentum transferred and the impact parameter of the collision, we can derive the average linear momentum imparted to the target nucleus in the case of each energy under study. The full momentum transferred would correspond to the momentum for the compound nucleus formation (CF). Consequently, the relative value of the transferred momentum p/p$_{CN}$ (where $p_{CN}$ is the full momentum transferred), containing the information on the initial reaction mechanism, can be obtained and equals for $^{197}$Au at 137.5 and 255.5 MeV 0.77$\pm$0.08 and 0.84$\pm$0.08, respectively. These values indicate that, in the intermediate energy range of projectile, the processes such fission, evaporation-spallation and intermediate fragment formation do not proceed solely via the compound nucleus formation. Other mechanisms are taking participation on the first step reaction.
We can say that at intermediate energy regime and high angular momentum, imparted to the system, the more probable processes proceed via surface interactions. On the other side, there are suggestions that there is no sharp boundary in the coexistence of the CF and ICF processes. Such coexistence is possible at both low and high energies just with different proportion. 

The analysis of the experimental data obtained in this work has shown the presence of a number of the isotopes close to the target what can be considered as a presence also of the CF process. The subsequent analysis of experimental data is needed to gain a better insight into the reaction dynamics involved at energies under study.

\section*{4 Conclusion}

The cross sections of the target fragments have been determined by the induced-activity techniques for the interaction of $^{197}$Au target with $^{11}$B-ions at the energies 137.5 and 255.5 MeV. Assuming a Gaussian charge distribution, the mass-yield distributions have been deduced. The mass-yield distribution of fission fragments at 137.5 MeV has been analyzed via the multimodal fission approach. The analysis has shown the two main fission modes (symmetric and asymmetric) to be determined by two distinct valleys in
the deformation potential energy, which had been theoretically predicted by Pashkevich.
The contribution of the asymmetric component to the total fission cross section is totally absent 255.5 MeV. This fact can manifest about gradually smearing the shell effects and vanishing completely at intermediate energy regime. The relative values of the linear momenta imparted to the targets, which contain the information on the initial reaction mechanism, were deduced using angular momenta and impact parameters of the interactions on the base of total reaction cross sections. These values indicated that, in the intermediate energy regime for pre-actinide targets, the interaction with heavy ions does not proceed solely via the compound nucleus formation. Other mechanisms are taking participation on the first step of reaction, and there are evidence of the coexistence both CF and ICF processes.

\medbreak

\newpage
\begin{center}
Table I. Cross section of the fission fragments formed by the reaction of 137.5 and 255.5 MeV
${}^{11}$B-ions with ${}^{197}$Au. Independent cross sections are indicated by (I); others are cumulative (C).
\end{center}
\begin{center}
\begin{tabular}{||c|c|c|c|c|c|c|c||}
\hline\hline
Element&Type& \multicolumn{2}{|c|}{Cross section, mb}&Element&Type& \multicolumn{2}{|c|}{Cross section, mb}\\
\cline{3-4}\cline{7-8}
& & 137.5 MeV& 255.5 MeV& & & 137.5 MeV& 255.5 MeV\\ \hline
\hline $^{7}$Be{}&{}I{}&9.8$\pm$1.1&18.5$\pm$2.3&$^{113}$Ag{}&{}C{}&19.60$\pm$01.96&21.20$\pm$2.12\\
\hline $^{22}$Na{}&{}C{}&2.5$\pm$0.4&4.7$\pm$0.5&$^{115m}$Cd{}&{}C{}&8.02$\pm$0.80&15.70$\pm$1.57\\
\hline $^{24}$Na{}&{}C{}&2.2$\pm$0.2&4.3$\pm$0.4&$^{117g}$Cd{}&{}C{}&3.16$\pm$0.32&3.80$\pm$0.40\\
\hline $^{28}$Mg{}&{}C{}&0.9$\pm$0.1&1.3$\pm$0.1&$^{117m}$Cd{}&{}C{}&3.78$\pm$0.40&4.90$\pm$0.49\\
\hline $^{34m}$Cl{}&{}I{}&0.4$\pm$0.05&0.7$\pm$0.1&$^{117g}$In{}&{}C{}&1.60$\pm$0.17&1.70$\pm$0.20\\
\hline $^{38}$S{}&{}I{}&0.15$\pm$0.02&0.33$\pm$0.6&$^{117m}$In{}&{}I{}&0.95$\pm$0.09&0.99$\pm$0.10\\
\hline $^{39}$Cl{}&{}C{}&0.11$\pm$0.02&0.25$\pm$0.03&$^{117m}$Sn{}&{}I{}&-&0.20$\pm$0.02\\    
\hline $^{41}$Ar{}&{}C{}&0.08$\pm$0.01&0.15$\pm$0.03&$^{120m}$Sb{}&{}I{}&1.83$\pm$0.18&2.80$\pm$0.28\\
\hline $^{42}$K{}&{}C{}&0.09$\pm$0.01&0.15$\pm$0.02&$^{122}$Sb{}&{}I{}&2.14$\pm$0.21&2.78$\pm$0.30\\
\hline $^{46}$Sc{}&{}I{}&0.09$\pm$0.01&0.12$\pm$0.02&$^{124}$Sb{}&{}I{}&2.30$\pm$0.23&3.60$\pm$0.36\\
\hline $^{54}$Mn{}&{}I{}&0.08$\pm$0.02&0.25$\pm$0.03&$^{126}$Sb{}&{}C{}&0.46$\pm$0.06&0.61$\pm$0.06\\
\hline $^{57}$Co{}&{}I{}&0.14$\pm$0.02&0.37$\pm$0.04&$^{126}$I{}&{}I{}&0.86$\pm$0.09&1.50$\pm$0.15\\
\hline $^{59}$Fe{}&{}C{}&0.30$\pm$0.03&0.51$\pm$0.05&$^{127}$Sb{}&{}C{}&0.59$\pm$0.06&0.76$\pm$0.08\\
\hline $^{62}$Zn{}&{}C{}&0.30$\pm$0.03&1.20$\pm$1.12&$^{130}$I{}&{}I{}&0.50$\pm$0.06&0.80$\pm$0.09\\
\hline $^{65}$Zn{}&{}C{}&0.55$\pm$0.06&1.37$\pm$0.14&$^{131}$Ba{}&{}C{}&0.44$\pm$0.05&0.76$\pm$0.09\\
\hline $^{67}$Ga{}&{}C{}&0.76$\pm$0.08&2.61$\pm$0.26&$^{133}$I{}&{}C{}&0.60$\pm$0.08&1.10$\pm$0.11\\
\hline $^{69}$Ge{}&{}C{}&0.11$\pm$0.01&0.89$\pm$0.09&$^{135}$I{}&{}C{}&0.15$\pm$0.03&0.80$\pm$0.10\\
\hline $^{72}$Zn{}&{}C{}&0.63$\pm$0.06&1.47$\pm$0.15&$^{136}$Cs{}&{}I{}&0.60$\pm$0.09&0.80$\pm$0.11\\
\hline $^{73}$Se{}&{}C{}&0.087$\pm$0.009&0.98$\pm$0.02&$^{140}$Ba{}&{}C{}&0.20$\pm$0.02&0.20$\pm$0.02\\
\hline $^{76}$As{}&{}I{}&2.01$\pm$0.20&2.87$\pm$0.30&$^{141}$La{}&{}C{}&0.15$\pm$0.02&0.17$\pm$0.02\\
\hline $^{77}$Ge{}&{}C{}&0.30$\pm$0.03&0.40$\pm$0.04&$^{175}$Ta{}&{}C{}&--&0.40$\pm$0.04\\
\hline $^{77}$Br{}&{}C{}&1.20$\pm$0.12&1.80$\pm$0.18&$^{177}$Ta{}&{}C{}&--&1.1$\pm$0.1\\
\hline $^{78}$As{}&{}C{}&3.42$\pm$0.34&5.53$\pm$0.55&$^{181}$Re{}&{}C{}&--&5.2$\pm$0.7\\
\hline $^{82}$Br{}&{}I{}&2.70$\pm$0.27&4.98$\pm$0.50&$^{182}$Os{}&{}C{}&--&12.7$\pm$2.0\\
\hline $^{83}$Rb{}&{}I{}&1.44$\pm$0.14&5.35$\pm$0.54&$^{183}$Os{}&{}C{}&--&9.9$\pm$1.5\\
\hline $^{83}$Sr{}&{}C{}&-&2.30$\pm$0.50&$^{185}$Ir{}&{}C{}&--&11.9$\pm$1.5\\
\hline $^{84}$Rb{}&{}I{}&5.47$\pm$0.60&8.41$\pm$0.93&$^{186}$Ir{}&{}C{}&--&77.9$\pm$11.7\\
\hline $^{86}$Rb{}&{}I{}&16.20$\pm$1.60&22.30$\pm$2.23&$^{186}$Pt{}&{}C{}&0.5$\pm$0.1&19.7$\pm$2.9\\
\hline\hline
\end{tabular}
\end{center}
\vspace{2cm}

\begin{center}
Table I. (Continued.)
\end{center}
\begin{center}
\begin{tabular}{||c|c|c|c|c|c|c|c||} \hline
\hline $^{87g}$Y{}&{}I{}&1.13$\pm$0.10&1.30$\pm$0.26&$^{187}$Pt{}&{}C{}&--&14.1$\pm$2.1\\
\hline $^{87m}$Y{}&{}C{}&1.48$\pm$0.15&2.20$\pm$0.22&$^{188}$Pt{}&{}C{}&0.88$\pm$0.12&28.6$\pm$4.4\\
\hline $^{89}$Zr{}&{}C{}&1.75$\pm$0.18&2.37$\pm$0.24&$^{189}$Pt{}&{}C{}&2.6$\pm$0.4&60.9$\pm$9.1\\
\hline $^{90m}$Y{}&{}I{}&17.80$\pm$1.80&20.20$\pm$2.02&$^{191}$Au{}&{}C{}&12.3$\pm$1.5&81.5$\pm$12.2\\
\hline $^{91}$Sr{}&{}C{}&28.11$\pm$3.00&34.10$\pm$3.41&$^{192}$Au{}&{}I{}&10.9$\pm$1.4&38.7$\pm$5.0\\
\hline $^{91m}$Y{}&{}I{}&3.30$\pm$3.00&4.58$\pm$0.50&$^{192}$Hg{}&{}C{}&13.7$\pm$3.0&49.4$\pm$7.2\\
\hline $^{92}$Sr{}&{}C{}&22.30$\pm$2.23&25.20$\pm$2.52&$^{193}$Au{}&{}I{}&9.7$\pm$1.2&27.6$\pm$3.1\\
\hline $^{92}$Y{}&{}I{}&4.80$\pm$0.48&9.30$\pm$0.93&$^{193(m+g)}$Hg{}&{}C&32.1$\pm$5.0&52.9$\pm$7.8\\
\hline $^{93}$Y{}&{}C{}&28.20$\pm$2.82&33.50$\pm$3.35&$^{194}$Au{}&{}I{}&31.2$\pm$4.0&37.8$\pm$5.0\\
\hline $^{95}$Zr{}&{}C{}&26.40$\pm$2.64&29.10$\pm$2.90&$^{195(m+g)}$Hg{}&{}C&69.6$\pm$10.4&71.1$\pm$10.7\\
\hline $^{95g}$Nb{}&{}I{}&1.30$\pm$0.13&2.80$\pm$0.28&$^{195}$Tl{}&{}C{}&--&40.1$\pm$6.0\\
\hline $^{95m}$Nb{}&{}I{}&1.18$\pm$0.12&2.20$\pm$0.22&$^{196(m+g)}$Au{}&{}I&204.1$\pm$30.0&112.8$\pm$16.9\\
\hline $^{96}$Nb{}&{}I{}&15.22$\pm$1.52&17.94$\pm$1.80&$^{196(m+g)}$Tl{}&{}C&163.7$\pm$24.5&22.1$\pm$3.2\\
\hline $^{97}$Zr{}&{}C{}&22.10$\pm$2.21&25.3$\pm$2.53&$^{197}$Hg{}&{}C{}&64.7$\pm$9.7&49.9$\pm$7.5\\
\hline $^{97}$Nb{}&{}I{}&9.22$\pm$1.11&16.14$\pm$1.64&$^{197}$Tl{}&{}C{}&231.8$\pm$34.0&13.1$\pm$2.0\\
\hline $^{99}$Mo{}&{}C{}&14.88$\pm$1.50&24.50$\pm$2.45&$^{198}$Au{}&{}I{}&9.0$\pm$1.3&5.2$\pm$0.8$^{198}$\\
\hline $^{99m}$Tc{}&{}I{}&1.85$\pm$0.19&2.78$\pm$0.28&$^{198(m+g)}$Tl{}&{}C{}&77.3$\pm$9.8&55.5$\pm$8.3\\
\hline $^{102m}$Rh{}&{}I{}&3.37$\pm$0.34&5.76$\pm$0.58&$^{198}$Pb{}&{}C{}&9.1$\pm$1.4&3.8$\pm$0.6\\
\hline $^{103}$Ru{}&{}C{}&7.82$\pm$0.78&15.20$\pm$1.52&$^{199}$Tl{}&{}C{}&38.0$\pm$5.7&8.7$\pm$1.3\\
\hline $^{105}$Ru{}&{}C{}&10.22$\pm$1.02&12.80$\pm$1.28&$^{199}$Pb{}&{}C{}&15.6$\pm$2.3&8.6$\pm$1.3\\
\hline $^{105}$Rh{}&{}I{}&2.40$\pm$0.24&2.50$\pm$0.25&$^{200}$Tl{}&{}I{}&10.35$\pm$1.55&0.7$\pm$0.1\\
\hline $^{110m}$Ag{}&{}I{}&4.20$\pm$0.42&8.61$\pm$0.86&$^{200}$Pb{}&{}C{}&8.2$\pm$1.2&4.17$\pm$0.62\\
\hline $^{111m}$Pd{}&{}I{}&7.70$\pm$0.77&13.57$\pm$1.36&$^{203}$Pb{}&{}C{}&1.7$\pm$0.2&1.16$\pm$0.17\\
\hline $^{111}$Ag{}&{}I{}&0.70$\pm$0.07&3.11$\pm$0.31&$^{204}$Pb{}&{}C{}&7.6$\pm$1.1&9.6$\pm$1.4\\
\hline $^{112}$Pd{}&{}C{}&17.20$\pm$1.72&19.07$\pm$1.91&$^{205}$Bi{}&{}C{}&0.50$\pm$0.07&0.89$\pm$0.13\\
\hline $^{112}$Ag{}&{}I{}&1.34$\pm$0.13&2.81$\pm$0.28&--&--&--&--\\
\hline\hline
\end{tabular}
\end{center}
\vspace{2cm}

\newpage
\medbreak
\begin{center}
Table II. Fitted values of the parameters in (4) and fission cross section $\sigma_{f}$.
\end{center}
\begin{center}
\begin{tabular}{||c|c|c|c|c||}
\hline\hline
Parameter& \multicolumn{2}{|c|}{Value}\\
\cline{2-3}
& 137.5 MeV& 255.5 MeV\\ \hline\hline
$K_{AS}$&1.77$\pm$0.02&--\\ \hline
$K'_{AS}$&1.77$\pm$0.02&--\\ \hline
$\sigma_{AS}$&2.8$\pm$0.1&--\\ \hline
$\sigma'_{AS}$&2.8$\pm$0.1&--\\ \hline
$D_{AS}$&40.0$\pm$1.6&--\\ \hline
$K_S$&119.7$\pm$11.2&308.75$\pm$15.1\\ \hline
$\sigma_{S}$&11.57$\pm$0.67&12.61$\pm$0.89\\ \hline
$\overline{A}_{S}$&99.0$\pm$0.5&97.0$\pm$0.7\\ \hline
$\nu_{T}$&10.0$\pm$1.5&14.0$\pm$2.1\\ \hline
$\sigma_{f}$ (mb)&601.77$\pm$90.3&785.0$\pm$118.0\\ \hline
$\sigma_{AS}$ (mb)&1.77$\pm$0.3&--\\ \hline
$\sigma_{S}$ (mb)&600.0$\pm$90.0&785.0$\pm$118.0\\ \hline
$\sigma_{tot}$ (mb)&1653.5$\pm$248.0&1748.0$\pm$262.0\\ \hline\hline
\end{tabular}
\end{center}
\vspace{2cm}

\newpage
\begin{figure*}[h!]
\includegraphics[width=16cm]{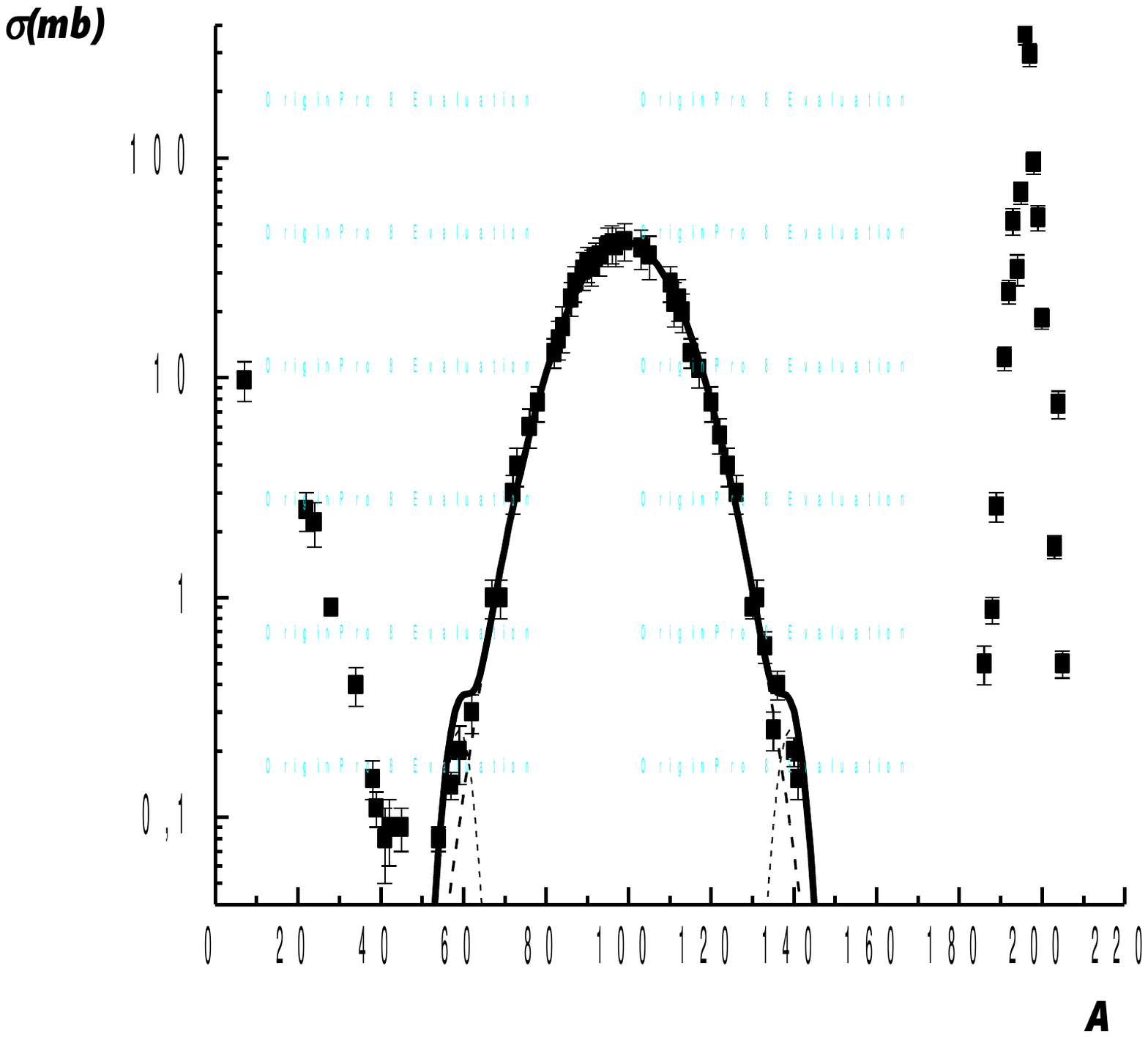}
\caption{\small Mass-yield distribution for the interaction of $^{197}$Au
with 137.5 MeV $^{11}$B ions. The dashed curves corresponds to the isobaric symmetric and asymmetric fission cross sections as a function of the mass of the fragments $A$. The black continuous curve corresponds to the total mass-yield, and experimental data are represented by the solid squares.}
\end{figure*}

\newpage
\begin{figure*}[h!]
\includegraphics[width=16cm]{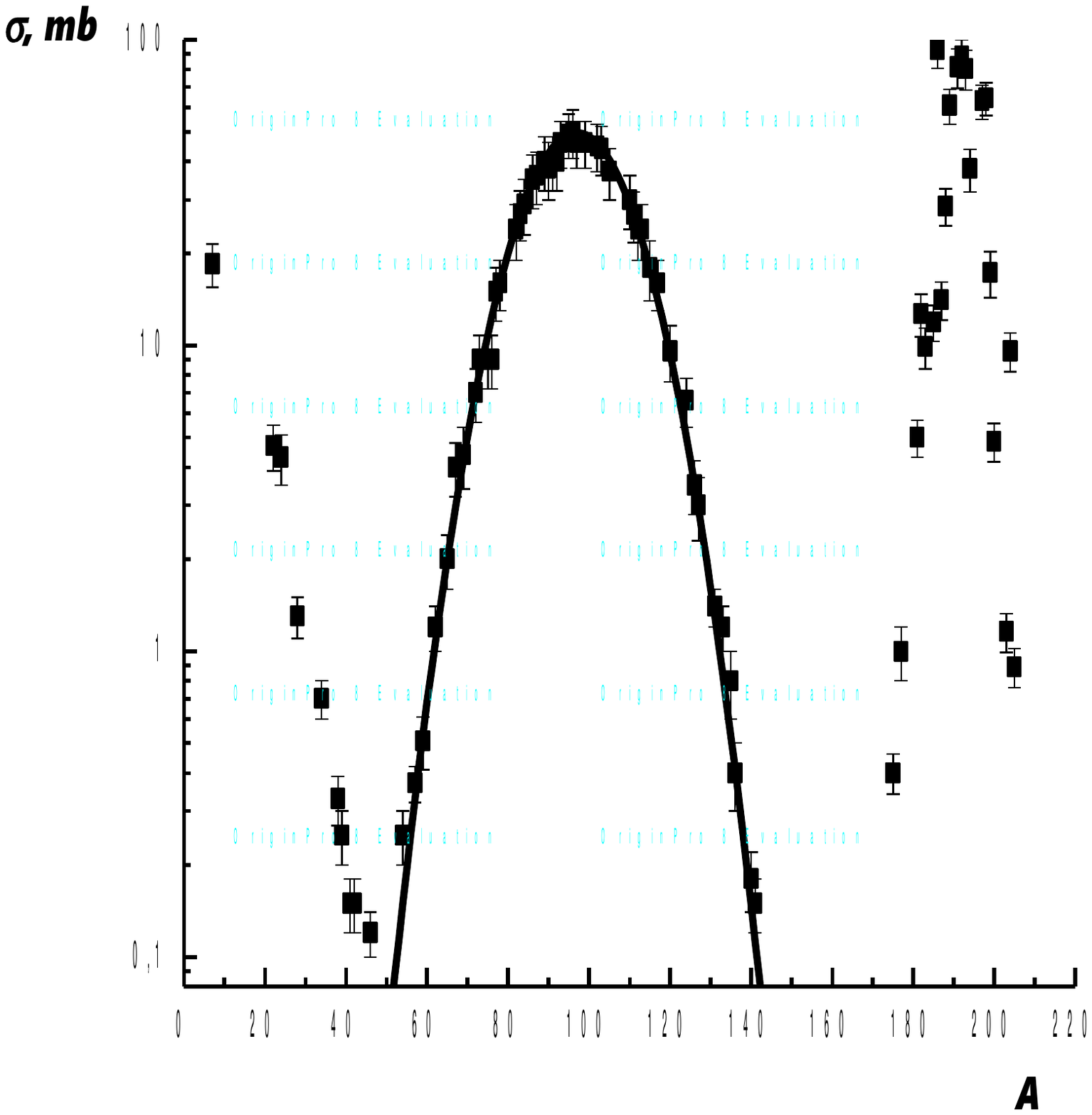}
\caption{\small Mass-yield distribution for the interaction of $^{197}$Au
with 255.5 MeV $^{11}$B ions. The dashed curves corresponds to the isobaric symmetric and asymmetric fission cross sections as a function of the mass of the fragments $A$. The black continuous curve corresponds to the total mass-yield, and experimental data are represented by the solid squares.}
\end{figure*}


\bigskip

\begin{thebibliography}{99}
\bibitem{Wilczynski} J. Wilczynski, Nucl. Phys. A $\bf216$, 386 (2073).{}
\bibitem{Gadioli1} E. Gadioli, G. F. Steyn,  Nucl. Phys. A $\bf708$, 391 (2002).{}
\bibitem{Gadioli2} E.Gadioli,  C. Birattari, M. Cavinato $\textit{et al.}$, Nucl. Phys. A $\bf641$,271 (1998).{}
\bibitem{Aguilera} E. F. Aguilera, J. J. Kolata Phys. Rev. C $\bf85$ (2012) 014603.{} 
\bibitem{Prokofiev} A. V. Prokofiev, Nucl.Instrum. Methods A $\bf463$, 557 (2001).{}
\bibitem{Buttkewitz} A. Buttkewitz, H. H. Duhm, F. Goldenbaum $\textit{et al.}$, Phys. Rev. C $\bf80$, 037603 (2009).{}
\bibitem{Gindler} J. Gindler $\textit{et al.}$, Nucl. Phys. A $\bf145$, 337 (1970).{} 
\bibitem{Lisman} F. L. Lisman $\textit{et al.}$, Phys. Rev. $\bf140$, B863 (1965).{}
\bibitem{Gaya} G. Karapetyan, EPJP $\bf130$, 180 (2015).{}
\bibitem{Kudo} H. Kudo, M. Maruyama, and M. Tanikawa $\textit{et al.}$, Phys. Rev. C $\bf57$, 178 (1998).{}
\bibitem{Branquihno} C. L. Branquihno and V. J. Robinson, J. Inorg. Nucl. Chem. $\bf39$, 921 (1977).{}
\bibitem{Gasques} L. R. Gasques, D. J. Hinde, M. Dasgupya $\textit{et al.}$, Phys. Rev. $\bf79$, 034605 (2009).{}
\bibitem{Eyal} Y. Eyal, K. Beg, D. Logan $\textit{et al.}$, Phys. Rev.C $\bf8$, 1109 (1973).{}
\bibitem{Yariv} Y. Yariv and Z. Fraenkel, Phys. Rev. C $\bf20$, 2227 (1979).{}
\bibitem{Gonnenwein} F. Gonnenwein, ``Mass, charge and kinetic energy of fission fragments'', The Nuclear Fission Process (C. Wagemans, Ed.), CRC Press, Boca Raton, USA (1991) 287.{}
\bibitem{Turkevich} A. Turkevich, J. B. Niday, Phys. Rev. $\bf84$, 52 (1951).{}
\bibitem{Pashkevich} V. V. Pashkevich, Nucl. Phys. A $\bf169$, 275 (1971).{}
\bibitem{Brosa}U. Brosa, $\textit{et al.}$, Phys. Rep. $\bf197$, 167 (1990).{}
\bibitem{Younes} W. Younes, J. A. Becker, L. A. Bernstein $\textit{et al.}$, Nuclear Physics
in the 21st Century: International Nuclear Physics Conference (INPC 2001) (AIP, New York, 2001) [AIP Conf. Proc. $\bf610$, 673
(2001)].{}
\bibitem{Duijvestijn} M. C. Duijvestijn, A.J. Koning, $\textit{et al.}$, Phys. Rev. C $\bf59$, 776 (1999).{}
\bibitem{Maslov} V. M. Maslov, Nucl. Phys. A $\bf717$, 3 (2003).{}
\bibitem{Sierk} A. J. Sierk $\textit{et al.}$, Phys. Rev. C $\bf33$, 2039(1986).{}
\bibitem{Capurro} O. A. Capurro, D. E. DiGregorio, S. Gil, $\textit{et al.}$, Phys. Rev. C $\bf55$, 766 (1997).{}
\bibitem{Bradt} H. L. Bradt and B. Peters, Phys. Rev. $\bf75$, 1779 (1949).{}
\end{thebibliography}
\end{document}